\documentclass[10ft,tightenlines,preprintnumbers,showpacs,
superscriptaddress,nofootinbib]{revtex4}
\usepackage{graphicx}
\usepackage{amssymb}
\usepackage{amsmath,mathrsfs,verbatim}
\usepackage{times}
\usepackage{latexsym}

\begin{document}

\title{Whitepaper on Super-weakly Interacting Massive Particles for Snowmass 2013}


\author{Jose A. R. Cembranos}
\affiliation{
Departamento de  F\'{\i}sica Te\'orica I,
  Universidad Complutense de Madrid, E-28040 Madrid, Spain}

\author{Manoj Kaplinghat}
\affiliation{Department of Physics and Astronomy,
 University of California, Irvine, CA 92697 USA}

\begin{abstract}
Super-weakly interacting massive particles produced in the late decays of weakly interacting massive particles (WIMPs) are generic in large regions of supersymmetric parameter space and other frameworks for physics beyond the standard model. If their masses are similar to that of the decaying WIMP, then they could naturally account for all of the cosmological dark matter abundance. Their astrophysical consequences and collider signatures are distinct and different from WIMP candidates. In particular, they could modify Big Bang Nucleosynthesis, distort the Cosmic Microwave Background, reduce galactic substructure and lower central densities of low-mass galaxies.
\end{abstract}

\maketitle


\section{INTRODUCTION}

The existence of dark matter (DM) is one of the major motivations for physics beyond
the standard model. However, our knowledge about DM arises purely from its gravitational effects. Collider, direct and indirect search null results only demand that this new form of matter couple weakly or super-weakly to the standard model (SM).

Particles with a weak-scale mass with renormalizable weak interactions to SM
particles have thermal relic density in the ballpark of the observed value. This fact motivates different searches for weakly interacting massive particles (WIMPs).
However, the complete stability of WIMPs is not mandatory, and late decays into super-weakly interacting massive particles (superWIMPs) are common in many theoretical frameworks. These new particles, with strongly suppressed couplings to the SM, can play the role of DM. In fact, if they are produced by the decay of WIMPs, their relic density could naturally be similar to that of WIMPs. In this sense, superWIMPs are as well-motivated dark matter candidates as the standard WIMPs.

\section{SuperWIMP Models}

Both WIMPs and superWIMPs emerge naturally in several well-motivated particle physics frameworks, such as
supersymmetry~\cite{SUSY}, universal extra dimensions~\cite{UED} and brane-worlds~\cite{BW1}.
However, the theoretical implications of superWIMPs are completely different from those of WIMPs.
We can illustrate this fact with $R$-parity conserving supersymmetry models, in which the lightest
supersymmetric particle (LSP) is completely stable. Within the WIMP scenario, the slepton LSP region
of the parameter space is excluded cosmologically.  In most part of the remaining allowed region,
the neutralino is the LSP.  Much of the neutralino LSP region is excluded because neutralinos are
overproduced. The situation is different within the superWIMP framework, where, for example,
the axino or the gravitino is the lightest supersymmetric particle (LSP). The regions of parameter space
where a slepton is the lightest SM superpartner are especially interesting, since late decays
to gravitinos can impact Big Bang nucleosynthesis (BBN) and possibly even resolve the anomalies associated with the primordial abundance of $^7$Li \cite{Feng:2003xh, BBN, boundstates}. Following the same argument, much of the region with neutralino as the next lightest supersymmetric partner (NLSP) is disfavored for the gravitino LSP case, since the hadronic neutralino decay typically destroys BBN successes.
On the other hand, regions excluded by overproduction within the classical WIMP framework, are the most interesting within the superWIMP scenario. In the standard case, where a WIMP decay produces one superWIMP, the abundance of the dark matter is reduced by the ratio of WIMP to superWIMP masses:

\begin{equation}
\Omega_\text{SWIMP}=\frac{m_\text{SWIMP}}{m_\text{WIMP}}\,\Omega_\text{WIMP}\,.
\label{omegas}
\end{equation}

Another possibility within this scenario is that of mixed warm and cold DM models.
A particularly interesting example arises when the axino is the LSP. The phenomenology
of such a superWIMP is characterized by a short WIMP lifetime compared to the gravitino LSP case. These models have received some attention in the last few years since they are less constrained by astrophysical observations \cite{axino}. Indeed, DM composed of a mixture of axion and axino has been claimed to be favoured in simple supersymmetric constructions \cite{mixed}.

\section{superWIMP signatures}

SuperWIMPs signatures have attracted a lot of interest from different points of view~\cite{Feng:2003xh, SWIMPs}.
This idea has provided new search strategies at colliders~\cite{CollSW,Hamaguchi:2004df,Feng:2004yi,Cembranos:2006gt}.
Depending on the charge of the decaying particle and its lifetime, superWIMP scenarios provide a rich
variety of exotic collider signals, such as displaced vertices, track kinks, tracks with nonvanishing impact parameters,
slow charged particles, and vanishing charged tracks \cite{Cembranos:2006gt}.
The decay lifetime could be months or even years. Given this, there have been different proposals to trap these particles (if charged) outside of the particle detector so that their decays can be analyzed and characterized~\cite{Feng:2004yi,Hamaguchi:2004df}.

On the other hand, the superWIMPs can leave their imprint on early universe cosmology. For example, the primordial element abundances may be modified due to energy injections from late time decays~\cite{BBN} and due to formation of new bound states with new meta-stable charged particles~\cite{boundstates}. These analyses can be used to constrain superWIMP models, but as we have commented above, in some regions of parameter space they may explain present inconsistencies in the Lithium abundance. This possibility can be corroborated. For instance, the heavy meta-stable charged particles could be produced in cosmic rays and detected with high energy neutrino telescopes or in sea water experiments \cite{Albuquerque:2003mi,Bi:2004ys}. In addition, late decays could also distort the Blackbody spectrum of the cosmic microwave background \cite{Yeung:2012ya} or be detected directly by studying cosmic ray spectra \cite{cosmic_rays}.

\begin{figure}[bt]
  \includegraphics[width=0.44\textwidth, clip=False, trim = 1mm 1mm 1mm 1mm]{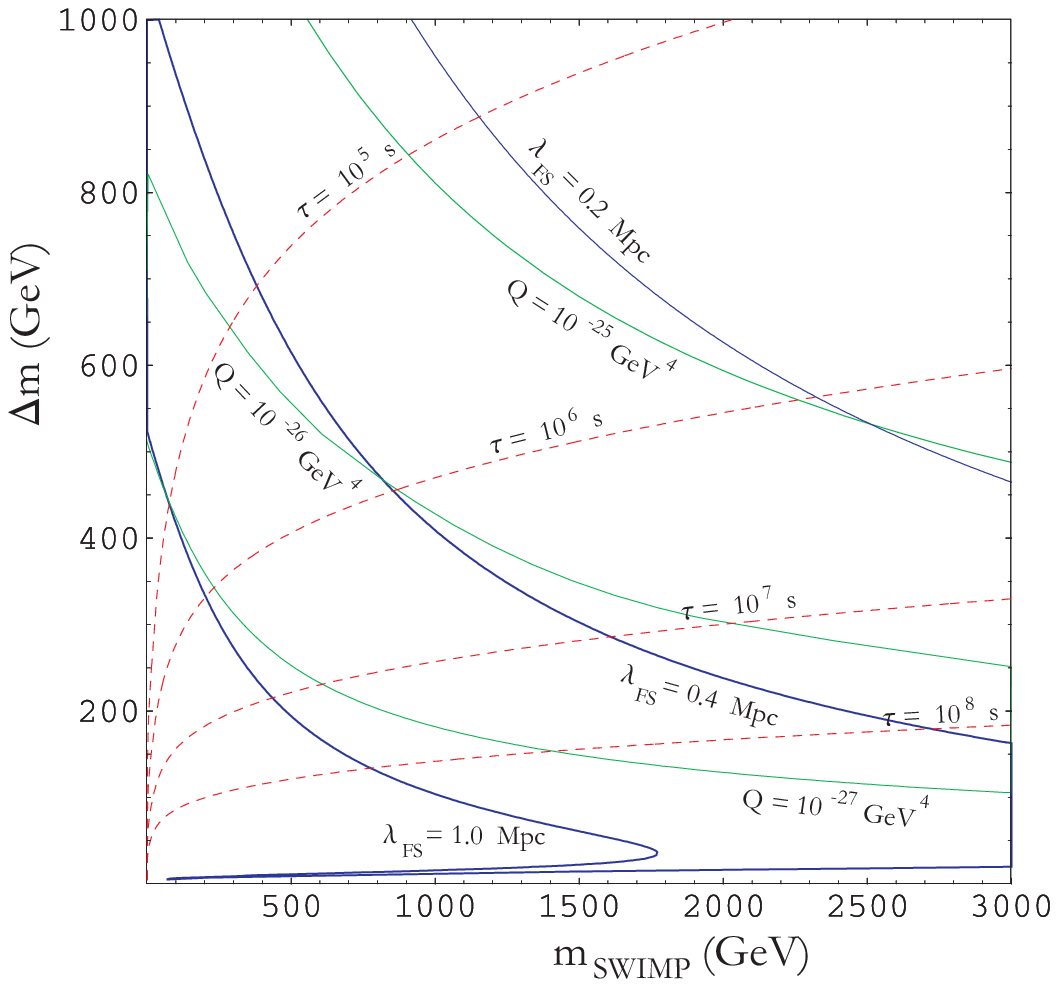}
  \includegraphics[width=0.45\textwidth, clip=True, trim = 10mm 10mm 10mm 25mm]{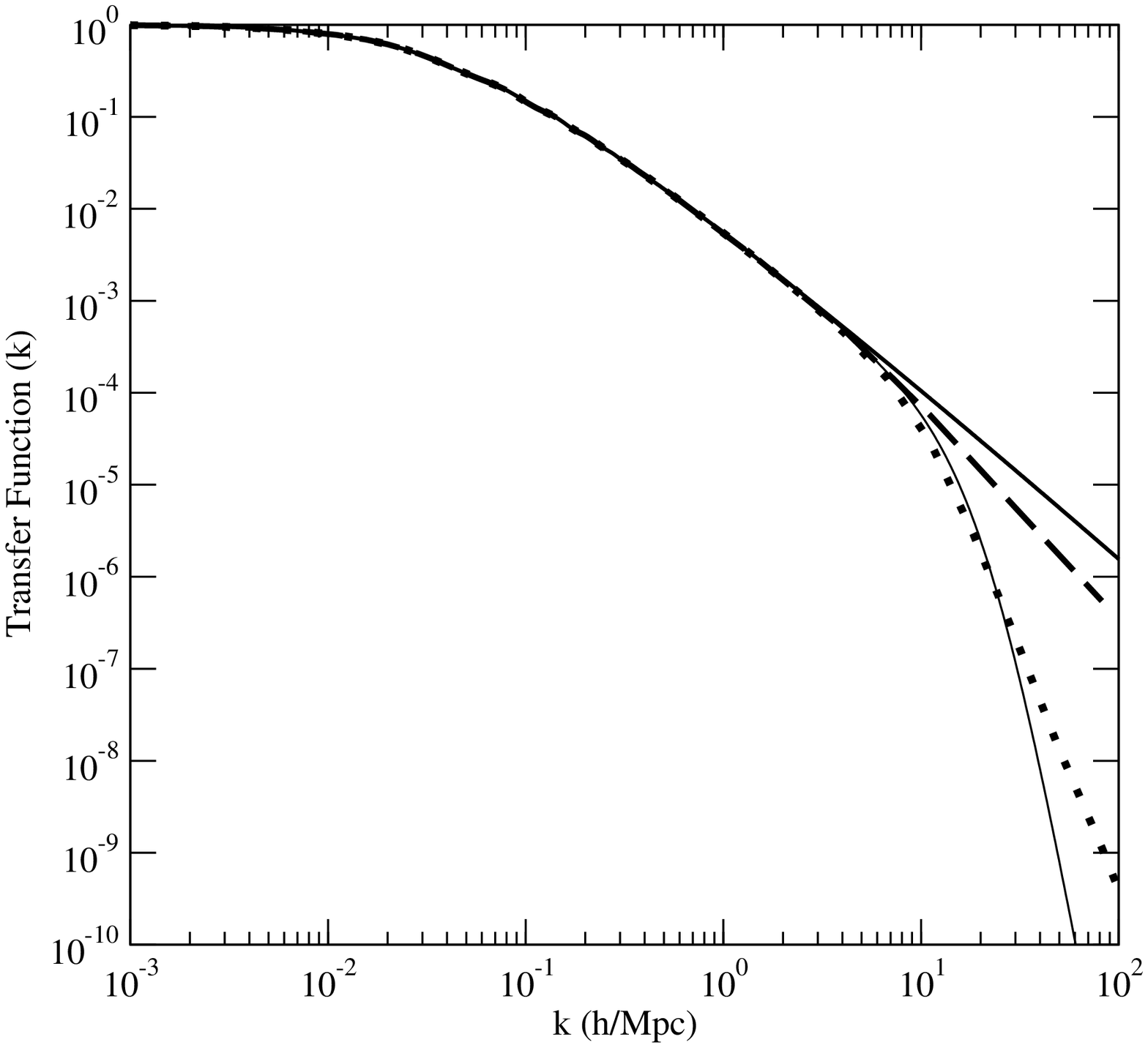}
\caption{{\bf Left}: SuperWIMP parameter space  $(m_\text{SWIMP}, \Delta m)$, where
$\Delta m \equiv m_\text{WIMP} - m_\text{SWIMP}$ for gravitino LSP (superWIMP)
and sneutrino NLSP (WIMP). $Q$ is the phase space density as defined by \cite{Hogan:2000bv}, $\lambda_\text{FS}$ is the free-streaming scale, and $\tau$ is the lifetime of the sneutrino. On the bottom-left corner, superWIMP DM behaves as
hot DM and it is excluded. On the top-right corner, it behaves as cold DM. Between both regimes, superWIMPs
work as a new type of warm DM that could reduce central densities and substructure in observable ways.
{\bf Right}: The curves show the power spectra for different values of $f$, the fraction of dark matter today that arises from decays as opposed to those produced during reheating (which would be cold dark matter). The solid curve shows the $f=0$ case (CDM). The dashed curve shows the $f=0.5$ case while the dotted curve shows the $f=1$ case. It is clear that the suppression on small scales is much reduced for the $f=0.5$ case. For comparison, we also plot (see thin solid curve) the transfer function for a 1 keV thermal Warm Dark Matter model.}
\label{sph}
\end{figure}

Despite their large masses, superWIMPs could behave (effectively) as warm dark matter (WDM) \cite{Cembranos:2005us,Kaplinghat:2005sy}. In fact, depending on the lifetime and the kinetic energy associated to the decay, they can work as hot, warm, cold or meta-cold DM. There are various puzzles in galaxy formation and one of the puzzles that has garnered much attention is the issue of the missing satellites \cite{Klypin:1999uc,Moore:1999nt}, essentially the question of how the thousands of subhalos in CDM simulations can be reconciled with the small number of Milky Way satellites discovered (about 20). In WDM models the formation of small mass halos is suppressed and hence it has been proposed as a solution for the missing satellites problem. Recent work has pointed out a new issue with Milky Way satellites -- the observed bright satellites are underdense in dark matter compared to the most massive subhalos of a Milky Way halo in CDM simulations \cite{BoylanKolchin:2011dk}. This is puzzling because it is expected that the most massive subhalos would host the bright satellites. Recent work has claimed that this issue is also solved in WDM models because subhalos have lower densities compared to their CDM counterparts \cite{Lovell:2011rd,Maccio':2012uh} due to the lack of power on small scales. The required WDM solution can be found within the context of superWIMP models by estimating the power spectrum cut-off scale~\cite{Sigurdson:2003vy,Cembranos:2005us,Kaplinghat:2005sy,Strigari:2006jf}. This is shown in more detail in Fig. \ref{sph} and we estimate that the region between the free-streaming scales of 0.2-0.4 Mpc could be the relevant WDM solution.

\section{CONCLUSIONS}

SuperWIMPs arise in well-motivated theoretical frameworks of beyond standard model physics.
These particles can inherit the relic abundance of WIMPs since they arise from the decay of a WIMP. The interactions of superWIMPs with standard model particles are strongly suppressed. However, there is a rich variety of distinctive signatures at colliders and in astrophysics and cosmology in this scenario.

\section{ACKNOWLEDGMENTS}

The work of JARC is supported by the Spanish MINECO projects numbers FIS2011-23000, FPA2011-27853-C02-01 and MULTIDARK CSD2009-00064
(Consolider-Ingenio 2010 Programme). MK is supported by NSF Grant PHY-1214648.

\end{document}